\documentclass[manuscript,screen, nonacm]{acmart}
\usepackage{xcolor} %approved
\usepackage{colortbl} %approved
\usepackage{multirow} %approved

\usepackage{booktabs,threeparttable}
\usepackage{multirow}

\newlength{\blocksize}
\newcolumntype{P}[1]{>{\centering\arraybackslash}p{#1}}

\AtBeginDocument{%
  \providecommand\BibTeX{{%
    \normalfont B\kern-0.5em{\scshape i\kern-0.25em b}\kern-0.8em\TeX}}}

\copyrightyear{2024}
\acmYear{2024}
\setcopyright{rightsretained}
\acmConference[CHI '24]{Proceedings of the CHI Conference on Human Factors in Computing Systems}{May 11--16, 2024}{Honolulu, HI, USA} \acmBooktitle{Proceedings of the CHI Conference on Human Factors in Computing Systems (CHI '24), May 11--16, 2024, Honolulu, HI, USA} \acmDOI{10.1145/3613904.3642118}
\acmISBN{979-8-4007-0330-0/24/05}

\begin{document}

\title{A Review of Virtual Reality Studies on Autonomous Vehicle--Pedestrian Interaction}


\author{Tram Thi Minh Tran}
\email{tram.tran@sydney.edu.au}
\orcid{0000-0002-4958-2465}
\affiliation{Design Lab, Sydney School of Architecture, Design and Planning,
  \institution{The University of Sydney}
  \city{Sydney}
  \state{NSW}
  \country{Australia}
}

\author{Callum Parker}
\email{callum.parker@sydney.edu.au}
\orcid{0000-0002-2173-9213}
\affiliation{Design Lab, Sydney School of Architecture, Design and Planning,
  \institution{The University of Sydney}
  \city{Sydney}
  \state{NSW}
  \country{Australia}
}

\author{Martin Tomitsch}
\email{Martin.Tomitsch@uts.edu.au}
\orcid{0000-0003-1998-2975}
\affiliation{Design Lab, Sydney School of Architecture, Design and Planning,
  \institution{The University of Sydney}
  \city{Sydney}
  \state{NSW}
  \country{Australia}
}

\renewcommand{\shortauthors}{Tran et al.}

\begin{abstract}
An increasing number of studies employ virtual reality (VR) to evaluate interactions between autonomous vehicles (AVs) and pedestrians. VR simulators are valued for their cost-effectiveness, flexibility in developing various traffic scenarios, safe conduct of user studies, and acceptable ecological validity. Reviewing the literature between 2010 and 2020, we found 31 empirical studies using VR as a testing apparatus for both implicit and explicit communication. By performing a systematic analysis, we identified current coverage of critical use cases, obtained a comprehensive account of factors influencing pedestrian behavior in simulated traffic scenarios, and assessed evaluation measures. Based on the findings, we present a set of recommendations for implementing VR pedestrian simulators and propose directions for future research.
\end{abstract}

% <concept_id>10003120.10003123.10011759</concept_id>
% <concept_desc>Human-centered computing~Empirical studies in interaction design</concept_desc>

% \ccsdesc[500]{Human-centered computing~Empirical studies in interaction design}

%% Keywords. 

\maketitle

\section{Introduction}
The driverless technology industry demonstrates enormous promise to revolutionize passenger transportation by increasing riders’ independence, comfort, and safety \cite{anderson2014autonomous}. Mass deployment of AVs, however, depends not only on technological progress and regulatory approval but also on public acceptance \cite{nordhoff2018acceptance}. Therefore, pedestrian interaction with AVs has been a research area of critical importance. Lacking any physical protections, pedestrians are the road users most vulnerable to fatalities and injuries in traffic collisions \cite{WHO2018}. Additionally, pedestrians are likely to be less familiar with self-driving technology compared to vehicle occupants \cite{robert2019future}. Practical applications aimed at solving AV--pedestrian interaction problems include pedestrian intention estimation and reasoning algorithms and human-centered design methods involving the overt communication of AVs \cite{rasouli2019autonomous}. The latter are the focus of this paper. Although pedestrians tend to rely on implicit vehicle movements to make crossing decisions \cite{moore2019case}, external human-machine interfaces (eHMIs) that utilize the vehicle’s exterior surface, its immediate surroundings, and wearable devices have been found to support safe and intuitive interactions \cite{dey2020taming}. 

While several methods exist to investigate AV--pedestrian interaction, the development of VR in the past decade has paved the way for the increasing use of immersive simulations. VR-based experiments obviate the following limitations of real-world testing: 1) Building fully functional prototypes and testing AVs in real traffic environments is expensive and time-consuming; 2) Regulations require the presence of a test driver to monitor and directly intervene if needed; and 3) Real-world study settings may entail considerable physical risks to the participants. Beyond addressing these limitations, VR simulators also offer great experimental control, enable researchers to reproduce other scholars’ work easily, and produce a multitude of behavioral data via the tracking system \cite{feng2020data}.

Popular simulation platforms, including screen-based setup, CAVE, and VR head-mounted displays (HMD), differ in their levels of complexity and immersion. Currently, researchers’ use of VR HMD is growing due to the technology’s high level of immersion, improved ergonomics, increased performance, and lower prices. While immersion is defined as a property of a VR system, presence is the subjective psychological response of a user experiencing that system \cite{slater1997framework}. Presence thus constitutes a major consideration in the development and validation of VR simulators. To a great extent, behavioral and emotional reactions elicited in the virtual environment (VE) were consistent with those in real-life situations \cite{deb2017efficacy}, and the average walking speed in VEs was found to match real-world norms \cite{deb2017efficacy}. Nevertheless, discrepancies between the real and virtual worlds remain an issue. Without a real danger of physical injury, participants were more inclined to exhibit risky behavior \cite{hollander2019investigating}. In addition, it was observed that participants overestimated vehicle speeds \cite{bhagavathula2018reality} and accepted a lower time-to-contact when crossing in VR \cite{feldstein2020road}. Technical limitations of VR such as a narrow field of view and relatively low display resolution might also impact the crossings, making it difficult for participants to spot oncoming vehicles and estimate distances\cite{mahadevan2019av}. Additionally, various side effects are associated with the use of VR, including discomfort, sickness, and other adverse after-effects \cite{Stanney2003}. While these shortcomings limit the generalization of obtained results, they neither invalidate the findings nor dismiss the usefulness of VR as an investigation method. 

% the perceived speed of the vehicles was significantly higher in both Unity and VUZE compared to the Smart Road \cite{bhagavathula2018reality}
% VR systems with non-surrounding displays (e.g., screen and projection) have a lower level of immersion compared to those with surrounding displays such as CAVE \cite{cruz1993surround} and head-mounted displays (HMD). Between CAVE and HMD, only the latter can substitute the participant’s real body parts with virtual ones (e.g. hand, lower body parts) and therefore achieves a higher level of immersion \cite{slater2016enhancing}. 

% Nevertheless, these differences in TTC must not be overvalued since the average absolute difference was a fraction of a second. 

% In order to use VR-based simulation effectively, researchers have been proposing different considerations to minimize the impact of the aforementioned challenges \cite{colley2019better}. 

% a real-world study setting could not generate the possibility that participants would get hit by the vehicle and would not create the sense of fear in them 

Several studies have analyzed the current use of VR simulators in pedestrian behavior research \cite{schneider2020virtually}, the development of AV safety \cite{nascimento2019role} and AV external communication research \cite{colley2019better}. In assessing 87 studies that employed pedestrian simulators, Schneider and Bengler \cite{schneider2020virtually} focused on four main characteristics: the research question, experimental task, technical setup, and participant sample. The review covered pedestrian behavior studies with and without AVs. Nascimento et al. \cite{nascimento2019role} examined the training of driving algorithms and the evaluation of user behavior in VEs. Since the research aimed to holistically examine the role of VR in enhancing the safety of automated driving technology, its analysis included only a small number of AV--pedestrian interaction studies. In a recent literature review by Colley et al. \cite{colley2019better}, the authors identified seven publications and preprints reporting the use of VR for AV external communication research. The simulators were analyzed based on their inherent advantages, such as customizability and objective data collection. Each of these works contributes to understanding VR potential as an evaluation method and provides considerations for future studies. 

Despite its contributions, the existing literature exhibits limitations. In particular, the increasing popularity of VR simulators in AV--pedestrian interaction research generates a vast number of traffic situations with diverse settings, which impedes effort to compare findings across studies and establish good implementation practices. It is, therefore,  imperative to review existing efforts and conduct a thorough analysis of investigated use cases and scenarios, as well as evaluation methods. The contribution of our work is threefold. Firstly, it summarizes the status quo. Secondly, it provides a set of considerations for researchers and industry practitioners. Thirdly, it highlights research gaps that are most pressing to further understanding AV--pedestrian interaction and any associated safety issues.

% Classification of use cases, as well as an extensive analysis of the simulation setup concerning different influencing factors of pedestrian behavior, have now become imperative to the transfer of knowledge. This paper contributes to the field by categorizing 31 VR-based studies on AV--pedestrian interaction, reviewing the scenario configuration, and analysing the evaluation measures. The contribution of our work is threefold. Firstly, it summarises the status quo. Secondly, it provides a set of considerations for researchers and industry practitioners. 

\section{Approach}

To perform an analysis of VR-based research on AV--pedestrian interaction, we adopted a structured approach based on the methodological framework by Arksey \& O’Malley \cite{Arksey2005}. 

\subsection{Research questions} This review is guided by the following questions:
\begin{enumerate}
  \item What are the main use cases of AV--pedestrian interaction being investigated in VR?
  \item How are the scenarios configured with respect to factors influencing pedestrian behavior? 
  \item What types of measures are used to evaluate AV--pedestrian interaction in VR?
\end{enumerate}

\subsection{Search strategies} 
\subsubsection{Data sources}
To identify relevant studies, we queried four research databases: ACM Digital Library, IEEE Xplore Digital Library, ScienceDirect, and Google Scholar. ACM Digital Library offers the most comprehensive collection of literature in computing and information technology. IEEE Xplore Digital Library provides access to high-quality technical publications in engineering and technology. ScienceDirect has a wide range of interdisciplinary research, while Google Scholar allows for a broad search across many disciplines and publishing formats. We restricted the search to include publications dated between 2010 and 2020 considering the increasing adoption of VR pedestrian simulators during this period \cite{schneider2020virtually}. The last search date was November 7, 2020.

\subsubsection{Keywords}
We selected four keywords that could represent the main concepts of our research topic, namely ``\textit{autonomous vehicles}", ``\textit{pedestrians}", ``\textit{interaction}" and ``\textit{virtual reality}". We also included synonyms and related words in making the search queries; however, abbreviations such as AV, VRU, HMI, VR have many different associated definitions and were not used to avoid irrelevant results. 

\subsubsection{Search procedure and results} We combined the keywords using AND/OR operators and performed the search within the title, abstract, keywords, and full-text of each article. After testing different keyword combinations, we selected the search query that yielded the highest number of results. The exact keywords and queries for each database can be found in Appendix A. The search yielded a total of 431 entries (ACM = 29, IEEE = 41, ScienceDirect = 19, Google Scholar = 342). We imported all research results to a spreadsheet to identify and remove 90 duplicate entries. One entry was not accessible to our academic institution. As a result, 341 publications remained. 

\subsection{Study selection} 
We chose conference proceedings and journal papers based on the following criteria:
\begin{itemize}
  \item Published as original full papers.
  \item Written in the English language.
  \item Contained empirical studies with evaluation results.
  \item Investigated AV--pedestrian interaction using VR as an evaluation method, with the AV--pedestrian communication being either explicit or implicit.
\end{itemize}

We excluded preprints, studies whose focus was to report the development of a VR simulator, and studies that offered little information to address the research questions. 

Since the number of studies found in our search was manageable, and the selection criteria were straightforward, one reviewer took charge of screening the publications. The process involved reading articles’ titles, abstracts, and full-texts to eliminate those that did not meet the selection criteria. A total of 31 publications was selected for the final analysis. A summary of these publications is available in Appendix B. 

\begin{figure*}
\centering
  \includegraphics[width=0.9\textwidth]{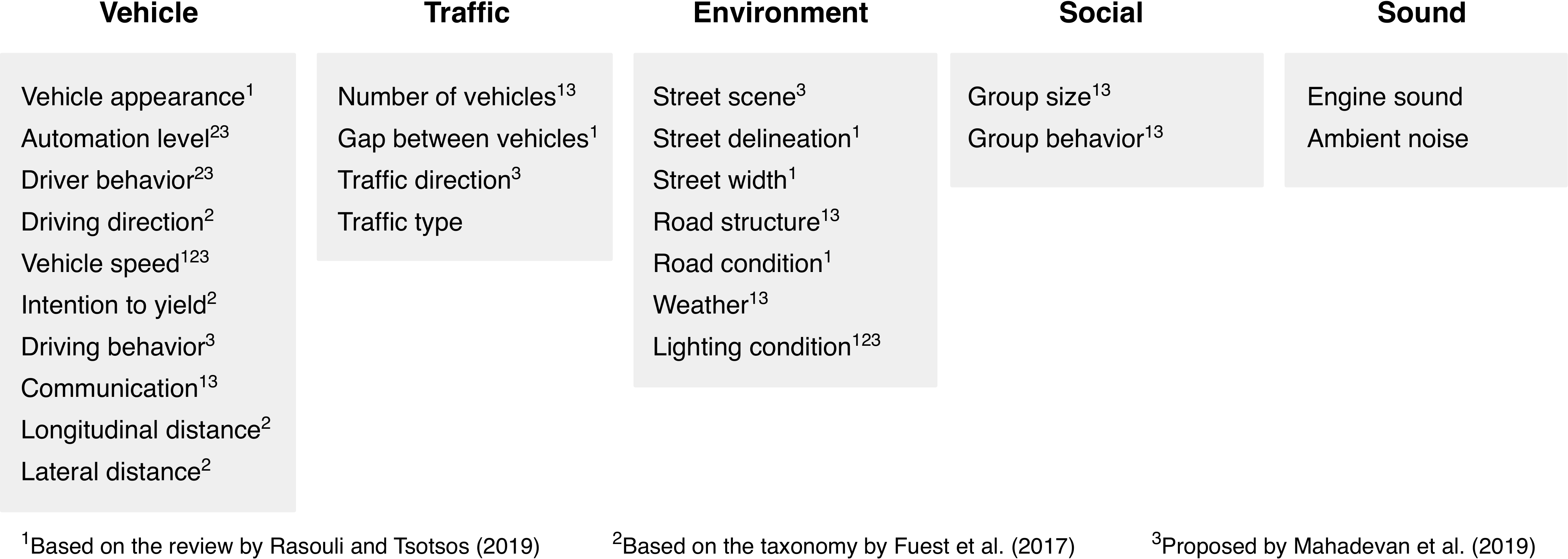}
  \caption{A summary of different factors influencing pedestrian experience and behavior.}
  \label{fig:scenario-config}
\end{figure*}

\subsection{Charting the data} 
Based on the research questions, we charted relevant information from each included publication and kept the records in a spreadsheet. The captured information was: author, year, title, VR system (platform), VR content type (computer-generated or 360-degree real-world video, contextualized or decontextualized), game engine, traffic scenario settings (see Fig. \ref{fig:scenario-config}), and evaluation measures. 

\section{Results}

This section presents the results from our analysis in a structure corresponding to the research questions. 

\subsection{Overview}
The analysis shows that immersive VR HMD was the system choice of all 31 reviewed studies. High-end headsets such as HTC Vive (22, 71\%) and Oculus Rift (3, 10\%) were favored over the low-budget phone-based VR (1, 3\%) since they offer powerful PC-tethered performance, fully-developed motion controls, and room-scale experience. Of note, five studies (16\%) did not specify the HMD model. In terms of content development, a large number of studies (30, 97\%) implemented contextualized scenarios. Nevertheless, one study (3\%) \cite{lee2019understanding} examined the comprehensibility of eHMIs in a decontextualized environment. Thirty studies (97\%) implemented computer-generated VR and one study (3\%) \cite{velasco2019studying} used a 360-degree real-world video. Concerning 30 studies whose simulations were made of 3D models, the VEs were consistently developed with Unity\footnote{https://unity.com/} (27, 87\%). Only one study (3\%) \cite{chen2020comparison} opted for the open-sourced jMonkeyEngine\footnote{https://jmonkeyengine.org/} which was made specifically for Java developers, and two studies (6\%) \cite{colley2020evaluating, jayaraman2018trust} did not mention the game engines used.

\subsection{Critical use cases and scenarios}

Various attempts have been made to define the most relevant use cases and scenarios of AV--pedestrian interaction. For example, Wilbrink et al. \cite{wilbrink2017interact} derived main use cases from workshops involving all project partners. In determining the relevance of each use case, they considered factors such as accident data, frequency of occurrence, the necessity of interaction, and cross-cultural differences. The outcome included two essential use cases concerning interactions in ambiguous situations, specifically at crossings without traffic lights and in car parks. Utilizing a different approach, Wang et al. \cite{wang2019safety} identified safety-critical scenarios from a video data set capturing pedestrian behavior at unmarked mid-block pedestrian crossings. The researchers subsequently classified scenarios based on the number of AVs and pedestrians in the interaction.

To highlight the extent to which existing research has investigated critical use cases and scenarios, we categorized 31 selected studies based on the right of way \cite{wilbrink2017interact} and the number of AVs and pedestrians in the interaction \cite{wang2019safety}.

\subsubsection{Right of way} 
All reviewed studies investigated unsignalized pedestrian crossings, except for one study (3\%) \cite{jayaraman2018trust} that examined the impact of both signalized and unsignalized crosswalks on pedestrians’ trust and certainty. Two studies (6\%) \cite{fuest2019should, weber2019crossing}, included ambiguous situations with undefined right of way, such as those in shared spaces and car parks. The absence of crossing facilities, such as crosswalks, indicated that AVs had the right of way in 17 studies (55\%), while their presence suggested that pedestrians had the right of way in 15 studies (48\%). Importantly, in countries such as Germany and the Netherlands, vehicles are obliged to yield to pedestrians at crosswalks; therefore, unmarked roads were chosen to ensure participants exercised caution \cite{ackermans2020effects, de2019external}. However, in places with different social norms and road rules, a crosswalk still represents an unsafe situation and requires participants to watch out for oncoming vehicles \cite{deb2020communicating}.

\subsubsection{The number of AVs and pedestrians in the interaction} 
Most studies (29, 94\%) involved one-to-one interactions. In making the crossing decision, participants only needed to observe one vehicle at a time. Fifteen studies (48\%) included multiple vehicles in traffic; however, the AV--pedestrian interaction in these experiments remained one-to-one. For example, in the study by Ackermans et al. \cite{ackermans2020effects}, 20 vehicles appeared in one experimental block, a new vehicle appeared only after the previous vehicle had been out of sight. Two studies (6\%) \cite{colley2020towards, mahadevan2019av} required the participant to interact with multiple vehicles while crossing. Specifically, the participant crossed a two-lane street where vehicles approached from both lanes. Only one study (3\%) \cite{mahadevan2019av} investigated the scenario of multiple vehicles - multiple pedestrians, and there was no instance where multiple pedestrians interacted with one vehicle.

% jayaraman

% \cite{deb2020communicating}
% In order to eliminate any visual blocking and/or distraction and to keep the experimental process controlled from environmental variations, no other vehicles were present in any of the scenarios while the participants were interacting with the trial car. The primary objective was to expose the participants to the features installed on the autonomous car and to develop the scenario in such a way that participants were able to make the crossing decision based only on their perception and comfort level with the features, not influenced by obstructions from other moving and/or stopped vehicles.

\subsection{Scenario configuration}

Even though there are a few instances of using 360-degree videos, the most common approach in VR simulation is to develop synthetic traffic scenarios. The process of creating a scenario involves the configuration of many different parameters, such as the number of vehicles and vehicle speed. Many of these parameters are also factors that influence the decision-making process of pedestrians when they cross the street. Rasouli and Tsotsos \cite{rasouli2019autonomous}, from their review of studies on pedestrian behavior, identified two groups of factors, namely pedestrian factors and environmental factors. Fuest et al. \cite{fuest2017taxonomy} derived from real traffic situations a set of attributes that influence communication between AVs and other road users (e.g., pedestrians, cyclists, and human drivers). Meanwhile, Mahadevan et al. \cite{mahadevan2019av}, through a preliminary design exercise and literature review, decided on 19 factors under four categories: pedestrians, vehicles, traffic and street characteristics, and interface prototypes in the development of a pedestrian–mixed-traffic simulator. To analyze the scenario configuration of selected VR studies, we synthesized factors relevant to AV–pedestrian interaction from the three studies mentioned above \cite{rasouli2019autonomous, mahadevan2019av, fuest2017taxonomy}, and grouped them into five categories using affinity diagramming. In the process, we added ``traffic type" to account for mixed traffic \cite{sae2021taxonomy} and ``sound," an important factor in navigation and decision making of the visually impaired \cite{koutsoklenis2011auditory} (see Fig. \ref{fig:scenario-config}).

This section presents the result of our analysis concerning: (1) which factors have been explored in the context of AV–pedestrian interaction and their influence on pedestrian experience and behavior; and (2) how the values of these factors are commonly specified in virtual traffic scenarios. We included the authors’ explanations, where given, for selecting a particular value (see Table \ref{tab:factor-table}). \\

\textbf{Vehicle factors}
% \cite{locken2019should,velasco2019studying,deb2019comparison,kooijman2019ehmis,de2019external,camara2020examining}. 

% The dominant vehicle type used in the reviewed studies was passenger cars (90\%), following by buses or shuttles (16\%), light trucks (6\%), heavy truck (3\%), and one-person pod (3\%) \cite{colley2020evaluating}. These vehicle selections were motivated by statistical data in which cars and light trucks accounted for most of the pedestrians’ fatalities \cite{deb2019comparison}, or the need to evaluate the interaction with different vehicle sizes.

\emph{Vehicle appearance - Type and size:} 
Depending on its type, a vehicle can be categorized as small, medium, or large. In this paper, small refers to single-seat or 2-seater vehicles; medium refers to 5-seater vehicles; and large refers to public buses, shuttles, and trucks. We created this coarse-grained classification of vehicle size based on the information from the reviewed studies rather than the actual dimensions of each vehicle model. The impact of vehicle sizes on AV--pedestrian interaction was investigated in three studies (10\%)\cite{de2019external,deb2019comparison,camara2020examining}. Clercq et al. \cite{de2019external} found that larger vehicles were perceived as less safe. However, the effect sizes were small for both yielding and non-yielding vehicles. Vehicle sizes were found to significantly affect the participant’s ability to comprehend eHMI messages \cite{deb2019comparison}. Specifically, the large vehicle size may have taken more time for the pedestrians to grasp the vehicle’s intentions. Only in one study \cite{camara2020examining}, did vehicle size not elicit any difference in pedestrian crossing behavior. However, it is worth noting that the number of participants in this study was low.

\emph{Vehicle appearance - Color:} 
White was the most common vehicle color, possibly because it makes the vehicle more noticeable in VR \cite{mahadevan2019av}. Nevertheless, whether a vehicle’s color may affect its visibility and hence influence pedestrian behavior was not explored in any reviewed studies. This factor might be relevant to consider in VR simulations, where light and color rendering is usually less realistic. 
% "LED visibility was dependent on light and also the car paint"
% \cite{mahadevan2019av} We fixed its color to white to make it easy to spot in VR

\emph{Vehicle appearance - Design:} 
A few studies tested friendly and futuristic vehicle models, such as Citroën C-Zero \cite{hollander2019investigating,hollander2019overtrust}, Waymo Firefly \cite{hudson2018pedestrian, deb2018investigating,deb2019comparison,deb2020communicating}, Mercedes-Benz F 015 \cite{locken2019should}, Easymile EZ10 and similar AV shuttles \cite{bockle2017sav2p, stadler2019tool, velasco2019studying}. Their futuristic designs might indicate the vehicles’ autonomous capability and help distinguish them from human-driven vehicles. Furthermore, conspicuous sensor systems mounted on top of an AV and eHMIs attached to its front made the vehicle’s autonomous nature more pronounced. Ackermans et al. \cite{ackermans2020effects} found that such a conspicuous appearance led to an increased willingness to cross by pedestrians who held a negative perception of AVs. 

% \cite{ackermans2020effects}
% "A conspicuous appearance of automated-driving capability had no effect for the sample as a whole, although it led to more efficient crossing decisions for those with a more negative attitude towards AVs"

\emph {Automation level:} 
By comparing pedestrian behavior when interacting with two types of vehicles – manually driven and fully autonomous – Chen et al. \cite{chen2020comparison} found a tendency of pedestrians to make crossing decisions based on their legacy strategy of gap acceptance, even if the vehicles they encountered were autonomous and equipped with eHMI displaying assistance information. Velasco et al. \cite{velasco2019studying} also reported no difference in pedestrians’ crossing intention; however, participants who recognized the automated-driving capability of the vehicle showed a higher level of trust in automation. It is worth noting that the higher trust scores somewhat contradicted the lower intention to cross; therefore, it was probable that the participants responded to the trust questionnaire while thinking about a future version of AVs instead of the currently available version \cite{velasco2019studying}. Only in one study (3\%) \cite{mahadevan2019av}, a semi-AV was included in mixed traffic, but without an eHMI, the vehicle’s automation level was ambiguous to the pedestrian participants. The study highlighted the potential problem of pedestrians’ inability to assess who is in charge of the vehicle’s operation and suggested that semi-AVs may need to communicate their driving statuses to other road users.

\emph {Driver behavior:} 
Two studies (6\%) \cite{deb2020communicating, hudson2018pedestrian} set out to identify the effect of AV operator statuses on pedestrian behavior. Studies by Deb et al. \cite{deb2020communicating} and Hudson et al. \cite{hudson2018pedestrian} reported no significant differences in external interface ratings between no-operator status and attentive-operator status. However, the presence of a distracted operator on a fully automated vehicle (SAE level 5, \cite{sae2021taxonomy}) negatively affected the participants’ perception of safety and lowered their ratings of the interface.

\emph{Driving direction:} There was no instance of backward-driving behavior even though a vehicle reversing into a car park was considered one of the must-have use cases because of the high number of accidents related to this situation \cite{wilbrink2017interact}. However, since the reversing maneuvers could be signaled by reversing lights and audible reversing alarms, many researchers considered this use case not as critical \cite{kass2020standardized}.

\emph{Vehicle speed:} 
The effect of vehicle speed on pedestrian crossing behavior was investigated in four studies (13\%) \cite{velasco2019studying, lee2019investigating, deb2019comparison,othersen2018designing}. When participants interacted with vehicles driving at a slower speed, they had a lower intention to cross \cite{velasco2019studying} and made fewer crossings \cite{lee2019investigating}. These findings could probably be explained by the fact that for the same time gap, vehicles traveling faster provided a larger distance gap \cite{lee2019investigating}. However, it is essential to note that a car traveling at higher speeds induced a lower safety margin \cite{lee2019investigating} and when it started to decelerate, pedestrians took longer to initiate the crossing \cite{othersen2018designing} and crossed less often \cite{lee2019investigating}. In a study by Deb et al. \cite{deb2019comparison} the impact of speed was examined in combination with distance. The results showed that while the adults exhibited normal behavior, the children made risky and hasty crossings when the AV was driving at high speed and the gap was narrow.

\emph{Intention to yield:} In two studies (6\%) \cite{chen2020comparison, stadler2019tool}, the vehicle would never yield its right of way. Participants were told to identify a safe gap to cross \cite{chen2020comparison} or were given sufficient time to cross a single lane of traffic \cite{stadler2019tool}. In most scenarios (22, 71\%), the AV could either yield to oncoming pedestrians or continue driving. The unpredictable behavior prompted participants to pay attention to vehicle kinematics or external car displays. In a smaller number of studies (5, 16\%), the AV decelerated in every trial. This setup was chosen because the vehicle had to comply with road traffic regulations \cite{fuest2019should}, the proposed eHMIs indicated safe crossing conditions only \cite{deb2018investigating,colley2020towards}, or different yielding behavior \cite{pillai2017virtual} and the impact of eHMIs in communicating yielding intent \cite{bockle2017sav2p} were the key parameters under inquiry in the research. However, a potential problem of this conservative behavior of AVs, as mentioned by Deb et al. \cite{deb2018investigating}, is that participants might expect the car to always stop for them, which might encourage them to step onto the road immediately instead of acting cautiously.

\emph{Driving behavior:} 
Typically, AVs were designed to maintain a certain speed or yield to pedestrians following a predetermined deceleration curve. However, in seven studies (23\%), different driving behaviors were included to explore how pedestrians inferred intentions from vehicle kinematics and their expectations of AV driving behavior. Results showed that an early deceleration was able to communicate an AV’s intention of giving way \cite{fuest2019should}, reduced the Crossing Initiation Time to Vehicle Stop (CIT\textsubscript{VS}) \cite{dietrich2020automated, dietrich2019implicit} and provided high levels of comfort to pedestrians \cite{pillai2017virtual}. Yielding early and slowly to pedestrians may also be a good approach to increase traffic efficiency in crossing situations since the AV might be able to accelerate again without making a complete stop \cite{dietrich2019implicit}. In contrast, late and abrupt braking was perceived as aggressive driving and decreased trust in AV technology \cite{jayaraman2018trust}. Atypical vehicle trajectories could even lead to mistrust \cite{schmidt2019hacking}. When interacting with the AV, participants related its driving behaviors to their own or other people’s actions \cite{pillai2017virtual}. It was found that people could derive social cues from different vehicle trajectories [50] and expected the AV to drive like a reasonable human driver [47]. Notably, in the study by Camara et al. \cite{camara2020examining}, participants interacted with a game-theoretic AV whose movement was adapted in real-time based on their positions. The authors reported that different participants displayed different preferences for AV algorithm parameters.

\emph {Communication:} 
The impact of eHMIs compared to no eHMI was explored in many studies (18, 58\%). In general, even though people still rely on vehicle kinematics to make crossing decisions \cite{chen2020comparison}, they expressed preferences for AVs to be equipped with external interfaces \cite{deb2020communicating}. Of note, older adults found the external interfaces more useful compared to their younger counterparts \cite{deb2020communicating}. Regarding subjective experience, eHMIs were found to provide pedestrians with a higher level of comfort \cite{mahadevan2019av, bockle2017sav2p}, perceived safety \cite{de2019external, bockle2017sav2p, velasco2019studying,chang2017eyes}, and trust \cite{colley2020evaluating}. When a vehicle explicitly communicated its intent to yield, this increased pedestrians’ crossing intention \cite{kooijman2019ehmis} and willingness to cross \cite{ackermans2020effects}. The users were able to make quicker crossing decisions \cite{chang2017eyes, stadler2019tool, hollander2019investigating} as well as initiating crossings earlier \cite{mahadevan2019av,dietrich2020automated,othersen2018designing}. Furthermore, a study by Deb et al. \cite{deb2018investigating} reported that the inclusion of an external interface led to a significant improvement in pedestrian’s receptiveness to AVs. Nevertheless, Weber et al. \cite{weber2019crossing} recommended the use of eHMIs in indicating safe crossing situations only, since a non-yielding intent might lead to more incorrect interpretations of the vehicle's intention.

\emph {Longitudinal distance:} This factor may influence objective safety and affect the timing of explicit communication, especially in the case of short distance \cite{fuest2017taxonomy}. In the study by Deb et al. \cite{deb2019comparison}, two distances were selected for every speed limit to present one risky situation (one that required hard braking) and one safe situation. The results showed that children took a longer time to cross the street during trials with wider gaps, whereas they rushed across the road during trials with higher speeds and narrower gaps. Koojiman et al. \cite{kooijman2019ehmis} reported that the eHMI had the strongest effect on pedestrians’ forward velocity when they crossed through a 20-meter gap (as opposed to a 30-meter gap).
% gap size (in meters) had the strongest effect on crossing intention, meaning that the distance is the most important factor affecting crossing intentions \cite{velasco2019studying}
% We look at scenarios in which pedestrians can see the car either shortly before crossing a road or in which it is difficult to interpret the car’s movements. In such situations, pedestrians cannot necessarily rely on guessing a vehicle’s behavior to take a decision. \cite{hollander2019investigating}

\emph{Lateral distance:} The setup for most studies (22, 71\%) was to have pedestrians standing at the curb of the pavement and the AV approaching from the nearest lane. In only three studies (10\%) \cite{hudson2018pedestrian,deb2018investigating,deb2020communicating}, the AV traveled from the farthest lane. \\

\textbf{Traffic factors}

\emph {Number of vehicles:} In about half of the reviewed studies (15, 48\%), participants encountered one vehicle in the scenario. However, many publications (15, 48\%) reported a setup in which multiple vehicles were driving past the participants’ location. In ten studies (32\%), participants could only cross between vehicles after one or several vehicles had passed \cite{chen2020comparison,dietrich2019implicit,dietrich2020automated,othersen2018designing,de2019external,lee2019investigating,deb2019comparison, deb2020communicating, hudson2018pedestrian,kooijman2019ehmis}. According to Chen et al. \cite{chen2020comparison}, by giving participants sufficient opportunity to prepare and adapt to the context, the setup could enhance the ecological perception of the following trial vehicle(s) and ensure that participants provide natural responses. 

% How about Jayaraman and Chang?

% However, it is important to note that having multiple vehicles in the scenario might increase cognitive load of the pedestrians. 

% However interaction 1-1: 
% - In investigating the influence of communication features, Deb et al. \cite{deb2020communicating} ensured participants were free from any visual distraction when interacting with the trial vehicle.
% In order to eliminate any visual blocking and/or distraction and to keep the experimental process controlled from environmental variations, no other vehicles were present in any of the scenarios while the participants were interacting with the trial car. The primary objective was to expose the participants to the features installed on the autonomous car and to develop the scenario in such a way that participants were able to make the crossing decision based only on their perception and comfort level with the features, not influenced by obstructions from other moving and/or stopped vehicles.

\emph {Gap between vehicles:} One critical aspect concerning the inclusion of multiple vehicles in the scenario is the (time or distance) gap between vehicles since this factor was found to influence pedestrians’ crossing decision. It is important to create a setup in which the participants cannot determine which vehicle will yield or provide a passable gap \cite{dietrich2020automated} so that they act as they would in a real-world traffic situation \cite{deb2020communicating}. Therefore, many studies varied the gap between vehicles in a fleet \cite{chen2020comparison,de2019external,lee2019investigating,othersen2018designing} or designed multiple experimental conditions with different gaps \cite{deb2019comparison,dietrich2020automated,dietrich2019implicit,kooijman2019ehmis}. In our analysis, three studies (10\%) \cite{deb2020communicating,hudson2018pedestrian, mahadevan2019av} reported impassable gaps between vehicles in the same convoy. These gaps were insufficient for the participants to make a safe crossing, which prompted them to look for deceleration cues or an explicit intention to yield in the approaching vehicles.

\emph {Traffic direction:} Most interaction scenarios involved traffic moving on one-way streets (27, 87\%). Two-way traffic was implemented in only three studies (10\%) \cite{hudson2018pedestrian, colley2020towards, deb2020communicating}. Koojiman et al. \cite{kooijman2019ehmis} suggested the testing of eHMIs in bidirectional traffic situations where participants have to divide their attention.

\emph {Traffic type:} The analysis found two studies (6\%) \cite{mahadevan2019av,chen2020comparison} examined AV--pedestrian interaction in mixed traffic. These efforts simulated two automation levels at a time. Chen et al. \cite{chen2020comparison} reported that a mixed-traffic environment of human-driven vehicles and fully automated vehicles did not affect pedestrian behavior. Mahadevan et al. \cite{mahadevan2019av} also did not find the influence of traffic composition on pedestrian crossing strategy to be statistically significant. \\

\textbf{Environment factors}

\emph {Street scene:} 
While most studies (29, 94\%) opted for a (Western) urban road environment, Mahadevan et al. \cite{mahadevan2019av} enabled two types of street scenes in their simulator: urban and rural. However, it remains unclear how these different environments might influence pedestrian behavior. Camara et al. \cite{camara2020examining} examined AV--pedestrian interaction using a narrow pathway in the park and a wide tarmac road, but the reported findings were still preliminary due to the low number of participants. Regarding the use of decontextualized VEs, while the absence of context might cause the eHMIs to be interpreted differently in the real world, it offers a fast and cost-effective method to investigate the intuitiveness of eHMIs’ colors and patterns in isolation from other factors \cite{lee2019understanding}. 

% In order to investigate the comprehension of the eHMI signal designs, Lee et al. \cite{lee2019understanding} simulated a decontextualized VE, in which operation of the eHMIs was decoupled from vehicle movement. 

\emph {Street delineation:} Traffic lights and zebra crossings are essential for the safe coexistence of people and vehicles. In four studies (13\%) \cite{velasco2019studying, jayaraman2018trust, fuest2019should, weber2019crossing}, the effect of these facilities was investigated in detail. Jayaraman et al. \cite{jayaraman2018trust} found that signalized crossings increased the trust in AVs and moderated the negative impact of aggressive driving behavior. At crossings, pedestrians also showed a higher intention to cross \cite{velasco2019studying}. Nevertheless, different rights of way did not elicit any differences in intention recognition time (IRTs) when pedestrians inferred an AV’s intent from its movement \cite{fuest2019should}. When the AVs are equipped with eHMIs, the effect of priority given to pedestrians’ correct interpretation, the certainty of choice, and IRTs varied depending on the cultural setting \cite{weber2019crossing}.

\emph {Street width:} A great majority of studies (18, 58\%) did not report street width, even though street width has been reported to impact the level of crossing risk \cite{rasouli2019autonomous}. Multi-lane streets are common in cities, and several study participants raised their concerns about whether the eHMIs displaying explicit instruction to act (e.g., a projected zebra crossing) would take into consideration other vehicles in other lanes \cite{locken2019should}.

% It may be dangerous to only display information for the lane ahead and not the other lanes that a pedestrian needs to cross.\cite{locken2019should}
% The analysis shows that one-lane streets measured 3.25-3.75 m and two-lane streets measured 4.86-8 m. The configuration was typically based on the country’s standards, for example, 3.5 m single lane road in the UK and Germany \cite{lee2019investigating,locken2019should}.

\emph {Road structure:} In two studies (6\%) \cite{fuest2019should, weber2019crossing}, shared space and a car park were selected to represent traffic situations with undefined right of way. The remaining studies investigated AV–pedestrian interaction at mid-block crossings (19, 61\%) or intersections (9, 29\%). Since intersections have a higher number of conflict points, we argue that this type of road structure will play an important role in studies that examine traffic scenarios with a scaled-up number of vehicles and pedestrians.

% with a more complex communication relationship between AVs and pedestrians. 

\emph {Weather, road conditions and lighting conditions:} 
One study (3\%) \cite{pillai2017virtual} investigated pedestrians’ understanding of AV driving behavior under two contrasting visibility conditions: a sunny day and night-time rain and fog. The findings revealed that participants felt less comfortable in adverse weather conditions, tended to act more cautiously, and relied more on audio cues to assess the situation. \\

\textbf{Social factors}

\emph {Group size:} 
This aspect refers to the number of pedestrians interacting with the AV in the scenario and does not include background people, such as those standing idly or chatting on the pavement. One study (3\%) \cite{mahadevan2019av} included in the traffic scenarios several AI-based virtual pedestrians of varying demographics (gender, ethnicity, and age) who would cross the street together with the study participants. 

\emph {Group behavior:} 
Mahadevan et al. [14] configured two crossing behaviors for the virtual pedestrians, including early crossers (crossing when the vehicles start to slow down) and timely crossers (crossing when the vehicles are almost at a complete stop). The findings indicated that the impact of group influence on behavior was not statistically significant.\\

% \begin{mdframed}[hidealllines=true,backgroundcolor=gray!20]
% \textbf{C4 - \textit{Creating a social atmosphere}}
% The rendering of a few people on the pavement and their background chatter \cite{colley2020towards} can bring liveliness to the simulated world and contribute to a more realistic experience [cite]. Studies should place these extras at a distance to avoid diverting participants’ attention, especially when users might react strongly to humanoid agents in VR \cite{hoggenmueller2021context} because of the potential uncanny valley effect \cite{mori2012uncanny}.
% \end{mdframed}

% \textbf{C4 - \textit{Being mindful of virtual people}}
% The presence of city dwellers was rarely included in the reviewed studies. However, the rendering of some background people can bring liveliness to the simulated world and contribute to a more realistic experience. Our suggestion is to place these characters at a distance from the participants to avoid confounding the results and diverting attention. Especially when users might react strongly to humanoid agents, which could lead to a potential uncanny valley effect \cite{mori2012uncanny}. 

% On the sidewalk, there are some people talking to each other \cite{colley2020towards}
% The simulation also provided typical background noise like chatting people and engine sounds.

% A user-study reveals positive effects of an increasing number of co-located social companions on the quality of experience of virtual worlds, i.e., on presence, possibility of interaction, and co-presence. 

\textbf{Sound factors}

\emph {Engine sound:} Audio cues from approaching vehicles are essential not only for people with full or partial losses of eye-sight but also for those who use smartphones while walking. Yet, only three studies (10\%) \cite{de2019external,colley2020towards,pillai2017virtual} mentioned the implementation of driving sounds. In one study (3\%) \cite{colley2020towards}, an Acoustic Vehicle Alerting System designed for electric vehicles was utilized. 

\emph {Ambient sound:} The urban soundscape (i.e., natural sounds and human-produced sounds) was described in nine studies (29\%). Notably, the sound effect of traffic lights was simulated in one study (3\%) \cite{colley2020towards}, which included visually impaired individuals as participants.\\

\begin{table*}
\caption{The impact of different factors on pedestrian experience and behavior}
\label{tab:factor-table}
\scriptsize
\begin{threeparttable}[t]
\begin{tabular}{p{0.14\textwidth}p{0.07\textwidth}p{0.24\textwidth}p{0.13\textwidth}p{0.07\textwidth}p{0.2\textwidth}}
\toprule
\emph{Factor} & \emph{Studies\tnote{1}} & \emph{Impact on pedestrians} & \emph{Values\tnote{2}} & \emph{\# Cases\tnote{3}} & \emph{Notes} \\\midrule
\textbf{Vehicle factors}\\\midrule

Appearance: type-size & \cite{de2019external, deb2019comparison}  & Perceived safety \cite{de2019external}  & Car (small, medium)  & 28 (90\%) & \multirow[t]{4}{4cm}{Cars and light trucks accounted for most of pedestrians’ fatalities \cite{deb2019comparison}  } \\ 
&\cite{camara2020examining} & Understanding of messages \cite{deb2019comparison} &Bus/Shuttle (large)  & 5 (16\%) &  \\ 
&&&Light truck (large)  & 2 (6\%) &  \\ 
&&&Heavy truck (large)  & 1 (3\%) &  \\ 
&&&1-person pod (small)  & 1 (3\%) &  \\ 
\arrayrulecolor{black!15}\midrule

Appearance: color &&& White/Silver  & 14 (45\%) & \multirow[t]{4}{4cm}{White could be used to make the vehicle easier to spot in VR \cite{mahadevan2019av}  } \\ 
&&&Other  & 7 (23\%) &  \\ 
&&&Black/Dark   & 6 (19\%) &  \\ 
&&&Red  & 5 (16\%) &  \\ 
&&&Unspecified  & 4 (13\%) &  \\  \midrule

Appearance: design & \cite{ackermans2020effects}  & Willingness to cross \cite{ackermans2020effects}& External interface  & 24 (77\%) & \multirow[t]{3}{4cm}{Sensor mounted on top of the vehicle helps to distinguish AVs from conventional vehicles \cite{chen2020comparison}  } \\ 
&&&Futuristic design  & 9 (29\%) &  \\ 
&&&Normal appearance  & 7 (23\%) &  \\ 
&&&Conspicuous sensor  & 3 (10\%) &  \\  \midrule

Automation level  & \cite{velasco2019studying, chen2020comparison}  & Trust \cite{velasco2019studying} & Fully autonomous  & 31 (100\%) & \\ 
&&& Manually-driven  & 3 (10\%) &  \\ 
&&&Semi-autonomous  & 1 (3\%) &  \\  \midrule

Driver behavior & \cite{deb2020communicating, hudson2018pedestrian} & Perceived safety \cite{deb2020communicating,hudson2018pedestrian}  & No driver  & 23 (74\%) &  \\ 
&&&Unspecified  & 6 (19\%) &  \\ 
&&&Attentive & 5 (16\%) & \\ 
&&&Distracted & 4 (13\%) &  \\   \midrule

Driving direction &&& Forward  & 29 (94\%) & \\ 
&&& Not applicable  & 2 (6\%) &  \\  \midrule

Vehicle speed & \cite{velasco2019studying,deb2019comparison} & Crossing intention \cite{velasco2019studying}  & 30--50 km/h  & 22 (71\%) & \multirow[t]{4}{4cm}{50 km/h - Singapore’s speed limit regulations for one lane roads\cite{stadler2019tool}} \\ 
& \cite{lee2019investigating,othersen2018designing}& Crossing time \cite{deb2019comparison} & Less than 30 km/h  & 4 (13\%) &  \\ 
&& Crossing decision \cite{lee2019investigating} & More than 50 km/h & 4 (13\%) &  \\ 
&& Safety margin \cite{lee2019investigating} & Unspecified  & 4 (13\%) &  \\  
&& CIT\textsubscript{VS} \cite{othersen2018designing} & 0 km/h & 2 (6\%) &  \\ \midrule

Intention to yield &&& Vary  & 22 (71\%) & \multirow[t]{4}{4cm}{A 5-sec gap acceptance was implemented so that participants could cross without the AV slowing down \cite{stadler2019tool}} \\ 
&&&Always yield  & 5 (16\%) &  \\ 
&&&Never yield   & 2 (6\%) &  \\ 
&&&Not applicable  & 2 (6\%) &  \\  \midrule

Driving behavior & \cite{jayaraman2018trust,fuest2019should}  & Intention recognition \cite{fuest2019should} & Unspecified  & 18 (58\%) & \multirow[t]{4}{4cm}{Selected deceleration rates reflected normal braking in previous research \cite{ackermans2020effects, pillai2017virtual} or were suggested by the UK Department for Transport \cite{lee2019investigating}  }  \\ 
&\cite{camara2020examining,pillai2017virtual} &CIT\textsubscript{VS} \cite{dietrich2020automated,dietrich2019implicit}& Specified  & 9 (29\%) &  \\ 
&\cite{dietrich2020automated,dietrich2019implicit,schmidt2019hacking}& Trust \cite{jayaraman2018trust,schmidt2019hacking}  & Not applicable  & 4 (13\%) &  \\
&& Comfort \cite{pillai2017virtual}&&&  \\ 
&&&&& \\ \midrule

Communication & \cite{hollander2019investigating, mahadevan2019av}  & Comfort \cite{mahadevan2019av,bockle2017sav2p} & Visual eHMI  & 23 (74\%) &  \\ 
&\cite{velasco2019studying,chen2020comparison,colley2020evaluating} & Perceived safety \cite{de2019external,chang2017eyes,bockle2017sav2p,velasco2019studying} & Auditory eHMI & 9 (29\%) &  \\ 
&\cite{weber2019crossing,ackermans2020effects,de2019external,deb2020communicating} & Trust \cite{colley2020evaluating}& Haptic eHMI  & 1 (3\%) &  \\
& \cite{hudson2018pedestrian,deb2018investigating}& Intention recognition \cite{weber2019crossing} & Implicit & 8 (26\%)&  \\
& \cite{bockle2017sav2p,stadler2019tool}& CIT\textsubscript{VS} \cite{dietrich2020automated,mahadevan2019av,othersen2018designing}&&&  \\
& \cite{dietrich2020automated,othersen2018designing} & Crossing intention \cite{kooijman2019ehmis}&&&  \\ 
& \cite{chang2017eyes,kooijman2019ehmis, gruenefeld2019vroad} & Willingness to cross \cite{ackermans2020effects} &&&  \\ 
&& Decision time \cite{chang2017eyes,stadler2019tool, hollander2019investigating}&&&  \\
&& AV receptivity \cite{deb2018investigating} &&&  \\ \midrule

Longitudinal distance & \cite{deb2019comparison,kooijman2019ehmis} & Crossing time \cite{deb2019comparison}  & More than 10 m  & 20 (65\%) &  \\ 
&& Forward velocity \cite{kooijman2019ehmis} & Unspecified  & 9 (29\%) &  \\ 
&&& Not applicable  & 2 (6\%) &  \\  
&&& 3--10 m & 1 (3\%) &  \\ 
\midrule

Lateral distance &&& Nearest lane  & 22 (71\%) & \multirow[t]{4}{4cm}{The AV approached from the farthest lane to avoid possible AV--pedestrian collision \cite{deb2018investigating} } \\ 
&&&Unspecified  & 4 (13\%) &  \\ 
&&&Farthest lane   & 3 (10\%) &  \\ 
&&&Not applicable  & 2 (6\%) &  \\
\arrayrulecolor{black}\midrule
\end{tabular}
\begin{tablenotes}
     \item[1] Studies included the factor as an independent variable.
     \item[2] \emph{Not applicable} refers to studies in which the factor does not exist by design (e.g., decontextualized scenario, single vehicle).
     \item[3] The total number of cases might be greater than the number of selected papers because a study might include multiple vehicles or multiple scenarios.
   \end{tablenotes}
    \end{threeparttable}%
\end{table*}

\begin{table*}
\scriptsize
\begin{tabular}{p{0.14\textwidth}p{0.07\textwidth}p{0.20\textwidth}p{0.14\textwidth}p{0.07\textwidth}p{0.23\textwidth}}
Table I (continued)\\
\toprule
\emph{Factor} & \emph{Studies} & \emph{Impact on pedestrians} & \emph{Values} & \emph{\# Cases} & \emph{Notes} \\\midrule
\textbf{Traffic factors}\\\midrule
Number of vehicles &&& One vehicle  & 15 (48\%) & \multirow[t]{4}{4.3cm}{The absence of other vehicles was to ensure pedestrians’ decision free from environmental distractions \cite{deb2018investigating} } \\ 
&&&Multiple vehicles  & 15 (48\%) &  \\ 
&&&Not applicable  & 1 (3\%) &  \\  
% \midrule
\arrayrulecolor{black!15}\midrule

Gap between vehicles &&& Not applicable  & 16 (52\%) & \\ 
&&&Specified  & 12 (39\%) &  \\ 
&&&Unspecified  & 3 (10\%) &  \\ \midrule

Traffic direction &&& One-way  & 27 (87\%) & \\ 
&&&Two-way  & 3 (10\%) &  \\ 
&&&Not applicable  & 1 (3\%) &  \\  \midrule

Traffic type & \cite{mahadevan2019av, chen2020comparison}  && Homogeneous  & 29 (94\%) & \\ 
&&&Mixed traffic  & 2 (6\%) &  \\ 
&&&Not applicable  & 1 (3\%) &  \\ \arrayrulecolor{black}\midrule

\textbf{Environment factors}\\ \arrayrulecolor{black}\midrule

Street scene & \cite{camara2020examining}  && Urban  & 29 (94\%) & \multirow[t]{3}{4cm}{An urban setting was selected because of the high rate of accidents between vehicles and pedestrians \cite{pillai2017virtual}} \\ 
&&&Tarmac road  & 1 (3\%) &  \\ 
&&&Park (garden)  & 1 (3\%) &  \\ 
&&&Not applicable  & 1 (3\%) &  \\ 
\arrayrulecolor{black!15}\midrule

Street delineation & \cite{velasco2019studying,jayaraman2018trust}  & Crossing intention \cite{velasco2019studying}& Unsignalized  & 30 (97\%) & \multirow[t]{4}{4cm}{In countries where vehicles are obliged to yield to pedestrians at crosswalks; unmarked roads were chosen to ensure participants exercise caution \cite{ackermans2020effects,de2019external} } \\ 
& \cite{fuest2019should,weber2019crossing}& Trust\cite{jayaraman2018trust}&Signalized  & 1 (3\%) &  \\ 
&&&Unmarked road  & 17 (55\%) &  \\
&&&Marked road & 15 (48\%) &  \\
&&&Not applicable  & 1 (3\%) &  \\ \midrule

Street width &&& Unspecified  & 18 (58\%) & \multirow[t]{4}{4cm}{The configuration was based on the country’s standards, e.g., 3.5 m single lane road in the UK \cite{lee2019investigating} and Germany \cite{locken2019should}}\\ 
&&&Two-lane  & 8 (26\%) &  \\ 
&&&One-lane  & 3 (10\%) &  \\ 
&&&Four-lane  & 1 (3\%) &  \\ 
&&&Not applicable  & 1 (3\%) &  \\  \midrule

Road structure &&& Mid-block  & 19 (61\%) & \\ 
&&&Intersection  & 9 (29\%) &  \\ 
&&&Parking/Shared space  & 2 (6\%) &  \\ 
&&&Unspecified  & 2 (6\%) &  \\ 
&&&Not applicable  & 1 (3\%) &  \\  \midrule

Road condition &&& Normal  & 30 (97\%) & \\ 
&&&Wet  & 1 (3\%) &  \\ 
&&&Not applicable  & 1 (3\%) &  \\  \midrule

Weather & \cite{pillai2017virtual}  & Comfort\cite{pillai2017virtual} & Normal  & 29 (94\%) & \multirow[t]{5}{4cm}{The extreme situation served to explore how pedestrians interact with approaching vehicles in a poor visibility condition\cite{pillai2017virtual}}  \\ 
&& Crossing decision \cite{pillai2017virtual}& Cloudy  & 1 (3\%) &  \\ 
&&&Rainy/Foggy  & 1 (3\%) &  \\ 
&&&Not applicable  & 1 (3\%) &  \\ \midrule

Lighting condition & \cite{pillai2017virtual}  & Comfort\cite{pillai2017virtual} & Photopic (daylight)  & 29 (94\%) & \multirow[t]{5}{4cm}{A dusk scenario was selected to ensure visibility of the projected display concepts \cite{nguyen2019designing}} \\ 
&& Crossing decision \cite{pillai2017virtual}& Mesopic (twilight)  & 1 (3\%) & \\ 
&&& Scotopic (night) & 1 (3\%) &  \\ 
&&&Not applicable  & 1 (3\%) &  \\ \arrayrulecolor{black}\midrule

\textbf{Social factors}\\ \arrayrulecolor{black}\midrule
Group size & \cite{mahadevan2019av}  && One person  & 30 (97\%) & \\ 
&&&Multiple pedestrians  & 1 (3\%) &  \\ 
&&&Not applicable  & 1 (3\%) &  \\ \arrayrulecolor{black!15}\midrule

Group behavior & \cite{mahadevan2019av}  && Not applicable  & 30 (97\%) & \\ 
&&&Early crosser  & 1 (3\%) &  \\ \arrayrulecolor{black}
&&&Timely crosser  & 1 (3\%) &  \\ \arrayrulecolor{black}\midrule

\textbf{Sound factors}\\ \arrayrulecolor{black}\midrule
Engine sound &&& Unspecified  & 26 (84\%) & \\ 
&&&Included  & 3 (10\%) &  \\ 
&&&Not included  & 1 (3\%) &  \\ 
&&&Not applicable  & 1 (3\%) &  \\ \arrayrulecolor{black!15}\midrule

Ambient sound &&& Unspecified  & 19 (61\%) & \multirow[t]{5}{4cm}{Background noise was included to increase the participants’ sense of presence \cite{hollander2019overtrust}} \\ 
&&&Included  & 9 (29\%) &  \\ 
&&&Not included  & 2 (6\%) &  \\ 
&&&Not applicable  & 1 (3\%) &  \\ 
\arrayrulecolor{black}\midrule
\end{tabular}
\end{table*}

\subsection{Evaluation measures}

Responding to the third research question, this section classifies the measures used to evaluate AV--pedestrian interaction (see Table \ref{tab:measurement}) and discusses prevalent evaluation methods in VR. 

\begin{table} [t]
\scriptsize
% \begin{tabular}{@{} *4l @{}} 
\caption{Evaluating AV--pedestrian interaction in VR}
\label{tab:measurement}
\begin{threeparttable}[t]
\begin{tabular}{p{0.13\textwidth}p{0.15\textwidth}p{0.1\textwidth}}
\toprule
\emph{Main categories} & \emph{Subcategories} & \emph{\# Cases\tnote{1}} \\\midrule
Measurement & Both & 26 (84\%)  \\
&Subjective only & 4 (13\%) \\
&Objective only & 1 (3\%) \\ \midrule
Measure & Questionnaire/Rating & 28 (90\%)\\
&HMD-logged data & 21 (68\%)\\
&Interview & 17 (55\%)\\
&Controller-input data & 7 (23\%)\\
&Video/VR recording & 5 (16\%)\\
&Direct observation & 2 (6\%)\\
&Motion suit & 1 (3\%)\\ \midrule
Experimental task & Cross the street & 23 (74\%) \\
&Indicate response & 8 (26\%) \\ \midrule
Evaluated aspect\tnote{2} & Crossing behavior & 24 (77\%) \\
& Comprehension & 12 (39\%) \\
& Perceived safety & 12 (39\%) \\
& Preference & 8 (26\%) \\
& Trust & 8 (26\%) \\
& Comfort/Affect & 5 (16\%) \\
& Mental workload & 4 (13\%) \\
\arrayrulecolor{black}\midrule
\end{tabular}
\begin{tablenotes}
    \item[1] The total number of cases might be greater than the number of selected papers because a study might employ various measures.
    \item[2] The list is not exhaustive. We do not include rarely used metrics (with 1--2 occurrences).
   \end{tablenotes}
    \end{threeparttable}%
\end{table}

\subsubsection{Experimental tasks}
In eight studies (26\%), participants indicated their responses verbally \cite{velasco2019studying} or using a controller \cite{chang2017eyes, fuest2019should,lee2019understanding,mahadevan2019av, ackermans2020effects,de2019external, weber2019crossing}. The remaining studies implemented naturalistic walking, which is essential for pedestrians to display more realistic responses and for researchers to extract behavioral data \cite{de2019external}. However, in addition to ample physical space, this experimental task requires a careful VE design to ensure participants’ safety and enhance spatial presence. For example, to minimize the risk of collision between participants and objects in the physical world, researchers using HMD could implement a virtual blue grid showing the limits of the walking area \cite{dietrich2020automated} or conceal physical barriers of the real world with inconspicuous virtual object \cite{hollander2019investigating,hollander2019overtrust}. In terms of embodiment, allowing the users to see their feet and hands might enable a full-body ownership illusion \cite{petkova2008if, kooijman2019ehmis}, whereas providing tactile feedback as participants step down from the sidewalk is expected to yield a more compelling sense of presence than does the current flat-floor setup \cite{kooijman2019ehmis}. 

Naturalistic walking is particularly valuable when researchers seek insights into pedestrian crossing behavior. If the objective is to evaluate the clarity of eHMIs only, an immersive high-fidelity setup may not be required \cite{kooijman2019ehmis}. Furthermore, HMDs can display either computer-generated content or 360-degree videos. While the former evokes more realistic physiological responses, the latter elicit better psychological responses despite its limited interactivity \cite{higuera2017psychological}. Thorough research into the application of 360-degree videos in AV--pedestrian interaction could be helpful for researchers who seek to improve their prototyping process.

\subsubsection{Evaluated aspects}
Various behavioral data were collected through video recordings, direct observations, and tracking of pedestrian positions and head movements. General reactions (e.g., hesitating before crossing) and objective metrics (e.g., waiting time) allowed the experimenters to assess the efficacy of the designed interaction and gain insights into pedestrians’ decision-making process. Due to studies employing a wide variety of metrics, the definition and calculation of seemingly similar metrics might nevertheless vary from one study to another. For example, while Locken et al. \cite{locken2019should} calculated the crossing time from the time the AV started braking to the time the pedestrian reached the opposite sidewalk, Deb et al. \cite{deb2019comparison} chose to start the timer when the participant initiated the crossing.

Pedestrians’ attitudes and experiences were generally evaluated with consistent instruments. For instance, Likert scales were primarily utilized to measure perceived safety and comfort, while the NASA-TLX \cite{Hart1988} and Self-Assessment Manikin \cite{bradley1994measuring} were reliably validated to assess workload and emotions, respectively. However, no consensus exists regarding the appropriate instruments to measure trust. While the simplicity of Likert scales [37], [48], [58] makes them easy to administer, they examine only a particular aspect of trust and, therefore, have limited validity. Standardized questionnaires can evaluate multi-dimensional constructs; however, different studies \cite{colley2020towards, colley2020evaluating,jayaraman2018trust, locken2019should, velasco2019studying} employed different trust questionnaires, resulting in limited comparability across results.

\section{Discussion}
Overall, the reviewed studies focused on facilitating one-to-one AV--pedestrian interaction in ambiguous traffic situations where pedestrians could not predict AV intent, i.e., in unsignalized crosswalks. After considering myriad factors, researchers have identified vehicle-related aspects as a major influence on pedestrian experience and behavior. External communication research accounted for many studies, but the efficacy of the design concepts has been restricted primarily to traffic scenarios with typical environmental conditions and uncomplicated communication relationships. With regard to evaluation measures, the high experimental control of VR allows rich behavioral data to be collected, but these objective measures need to be obtained more consistently. 

In this section, we reflect on the results and discuss research gaps and avenues for future research. Additionally, we propose relevant considerations for examining AV--pedestrian interaction in VR, the need for standardization and benchmarking across VR-based studies, and the limitations of this review.

\subsection {Research gaps}
The great potential of VR pedestrian simulators lies in their ability to develop complex and high-risk traffic scenarios that are otherwise difficult to realize via real-world testing. In addition, VR enables the prototyping of design solutions that involve drastic changes in urban infrastructure (e.g., smart roads). Therefore, in the following parts, we highlight research gaps in AV--pedestrian interaction, which are crucial to address and can be explored with a virtual test-bed.

\subsubsection{Scalability}  
Many researchers have expressed their concerns about: (1) the possibility of external displays to impose high cognitive load onto pedestrians \cite{mahadevan2019av,robert2019future, moore2019case}; and (2) the need to select an optimal communication strategy for AVs to avoid misinterpretation when pedestrians with different rights of way are within the vehicle’s proximity \cite{dey2020taming, robert2019future}. These scalability-associated problems are, by no means, exhaustive and they are far from being resolved. At the early stage of AV--pedestrian interaction research, a majority of the studies have focused on single pedestrian and single AV scenarios, while very few have attempted scalability testing \cite{colley2020unveiling}. Nevertheless, as the number of eHMI design concepts introduced in academia and industry continues to increase \cite{dey2020taming}, the capability of eHMIs to cope with complex traffic situations in which multiple AVs and pedestrians cross paths has now become one of the most critical aspects determining its prevalence.

\subsubsection{Mixed traffic}  
Mixed traffic, consisting of vehicles with different levels of automation \cite{sae2021taxonomy}, requires further attention for several reasons. First, pedestrians might need to gauge the intention of various vehicle types (i.e., conventional vehicles, semi-AVs, AVs) and operator statuses (i.e., no driver, attentive driver, distracted driver), which can be mentally demanding. Second, some AV--pedestrian communication solutions, such as the ``omniscient narrator" concept in which one vehicle communicates on behalf of other vehicles \cite{colley2020towards}, seem improbable in mixed traffic. However, mixed traffic research remains relatively underexplored in the literature. In our review, only two studies \cite{mahadevan2019av, chen2020comparison} examined the impact of mixed traffic on pedestrian experience and behavior, and these efforts were limited to simulating two automation levels at a time.

 \subsubsection{Environmental conditions}  
To maximize the potential impacts of AVs on pedestrian safety, car manufacturers continue to develop AVs’ capabilities to drive in inclement weather and at night. By contrast, VR studies of AV--pedestrian interaction have mostly been conducted in temperate weather conditions during daytime hours. By including varying environmental conditions, future studies could generate additional design requirements and provide insights for developing more holistic and contextual solutions. Furthermore, pedestrians have been found to rely mainly on vehicle motion patterns in making crossing decisions \cite{moore2019case}. Hence, scenarios where vehicle movement becomes difficult to observe will yield more insights into the relevance of eHMI solutions.

\subsubsection{Vehicle behavior in VR}  
Vehicles can be controlled by human participants in coupled or distributed simulators \cite{Bazilinskyy2020, sadraei2020vehicle}. In our analysis, however, all selected studies utilized the approach of programming vehicle behavior in VR. Through ray casting, a vehicle can detect any objects ahead of it and respond accordingly \cite{dalipi2020vr}. Nguyen et al. \cite{nguyen2019designing}, for instance, implemented a method to determine a vehicle’s driving behavior based on the presence of a human object in its visual field. To a considerable extent, these conditional instructions enable effective simulation of vehicle behavior; however, their simplicity does not allow real-time, back-and-forth interactions between AVs and pedestrians. Out of 31 reviewed studies, only one \cite{camara2020examining} utilized a game-theoretic model to continuously adapt the vehicle's driving behavior to participants' movement. Findings from this study suggested that mathematical models determining AV driving behavior could learn from real-world human inputs while adding value to VR pedestrian simulators. By allowing ongoing dynamic interaction between AVs and pedestrians, future research could investigate the negotiating process common among road users today while also enhancing simulation realism and participant's sense of presence.

% In our opinion, mathematical models employed widely to produce realistic traffic in driving simulators could also hold value for pedestrian simulators. Allowing an ongoing and dynamic interaction between AVs and pedestrians could potentially enhance the realism of the simulation and the participant’s sense of presence while also accurately reflecting the negotiating process common among road users today.

% Mathematical models employed widely to produce realistic traffic in driving simulators could also hold value for pedestrian simulators. 

\subsection {Considerations}
Based on our analysis of factors, we present five considerations to synthesize the observations from the reviewed studies in the form of actionable key takeaways. These considerations should be taken into account when developing future AV--pedestrian simulation studies.

\subsubsection{Simulating human-like driving behavior}
Vehicle kinematics remains the key contributing factor to pedestrian experience and behavior. In many studies, people stated their preferences for human-like vehicle driving behaviors \cite{pillai2017virtual} that conform to social expectations \cite{schmidt2019hacking}; therefore, it is essential for external communication research to replicate these behaviors accurately. Several studies have referred to local regulations and guidelines on speed limits \cite{stadler2019tool, velasco2019studying} and stopping distances \cite{lee2019investigating}. We suggest the use of naturalistic driving data \cite{ackermans2020effects, pillai2017virtual} or a tested driving profile that takes into consideration both the traffic context and pedestrians’ perception of safety.

\subsubsection{Developing appropriate traffic complexity}
To enable participants to sustain their attention on the trial AV and make crossing decisions based solely on the installed communication features or vehicle kinematics, studies should consider eliminating visual distractions that might be caused by other vehicles \cite{deb2020communicating}. This could be achieved by excluding ambient traffic \cite{deb2018investigating, hollander2019overtrust} or ensuring that the study participants interact with the trial car only when other vehicles are no longer present in the scene \cite{deb2020communicating, ackermans2020effects}. If, however, the study objective is to explore the efficacy of the design solutions in a realistic environment, including other moving traffic will introduce additional visual and auditory stimuli to the scenario.

\subsubsection{Considering familiar elements and traffic cultures}
The development of a familiar VE (e.g., an urban downtown scene \cite{deb2018investigating}, an inner-city road in Munich \cite{dietrich2020automated}, or a well-known suburb \cite{nguyen2019designing}) could increase study participants’ level of comfort \cite{nguyen2019designing}. Similar to a VR familiarization phase \cite{deb2020communicating}, these VEs might help reduce potential distractions caused by the novelty effect. Aside from the physical aspects, it is important to attend to the local traffic norms in designing traffic scenarios. For instance, an unsignalized crosswalk can represent either a safe \cite{ackermans2020effects, de2019external} or unsafe situation \cite{deb2020communicating} depending on cultures.

\subsubsection{Creating a social atmosphere}
Simply rendering a few people and their background chatter within a scene \cite{colley2020towards} can bring liveliness to the simulated world and contribute to a more realistic experience \cite{schuemie2001research}. These additions should be placed at a distance from the participant in a study's virtual environment to avoid diverting their attention, especially when users might react strongly to humanoid agents in VR \cite{hoggenmuller2021context} because of the potential uncanny valley effect \cite{mori2012uncanny}.

\subsubsection{Utilizing background noise} Including an ambient soundscape in the simulation was found to heighten the simulation’s realism and users’ perceived presence \cite{kern2020audio}. Therefore, studies should reproduce sound effects spatially in VR and consider suppressing the sounds of the actual environment with noise-canceling headphones \cite{stadler2019tool, othersen2018designing}. To validate the effect of auditory cues in a realistic implementation, studies must simulate an average urban noise level (e.g., 52 dB \cite{deb2020communicating}). However, if the simulated scenario features dense urban areas, the noise generated by road traffic and other street activity could become excessive and obscure any auditory signals intended for pedestrians \cite{mahadevan2019av}.

\subsection{Standardization and benchmarking}
Published studies have not always reported the configuration of VR scenarios in detail. Indeed, several factors, such as street width or vehicle slowdown characteristics, remained unspecified in many studies. The limited availability of vehicle driving data could pose a challenge for researchers trying to replicate studies. This is of particular importance as vehicle kinematics have been identified as essential cues on which pedestrians rely when making crossing decisions.

Except for a handful of VR pedestrian simulators that have been developed with a high level of customization \cite{deb2017efficacy,mahadevan2019av}, the majority of simulators have been implemented for specific study objectives. Consequently, these simulators have featured a limited number of traffic scenarios, eHMI design concepts, and vehicle trajectories. Furthermore, only a few of the published studies have made their simulator source codes publicly accessible. Only recently, Dalipi et al. \cite{dalipi2020vr} created a VR platform intended as an open-source benchmark that allows users to experience different interaction design concepts. We argue that making simulators open-source could also facilitate researchers’ efforts to compare different concepts.

Another significant aspect that needs standardization is the measures employed to assess AV--pedestrian interaction. While definitions and guidelines exist to determine driver performance in the field and driving simulators \cite{sae2015operational}, there are no established measures to evaluate pedestrian behavior and subjective experience when interacting with AVs. 

\subsection{Limitations} 
We cannot claim to have included all relevant studies. Indeed, studies could have been omitted due to our choice of keywords. However, it is intended that the identified factors and considerations are continually refined and built upon as the research field of AV--pedestrian interaction expands. Furthermore, not all reviewed papers included a colored figure or a video depicting the simulated scenario, making it challenging to retrieve all needed information for the analysis.

\section{Conclusion}
In this paper, we presented a comprehensive review of 31 VR-based studies on AV--pedestrian interaction. To uncover the focus of existing research efforts, traffic scenarios were classified based on the right of way and the number of pedestrians and AVs in the interaction. The analysis provided a detailed account of different factors influencing pedestrian experience and behavior and a set of considerations to accompany the development of VR pedestrian simulators. Additionally, we reported prevalent evaluation measures and highlighted research gaps that provide avenues for future research in this domain.

\begin{acks}
This research is supported by an Australian Government Research Training Program (RTP) Scholarship and through the ARC Discovery Project DP200102604, Trust and Safety in Autonomous Mobility Systems: A Human-centred Approach.
\end{acks}

\bibliographystyle{ACM-Reference-Format}
\bibliography{ACM/references}

\end{document}